\documentclass[12pt]{article}
\usepackage{physics}
\usepackage{epsf}
\usepackage{aas_macros}
\usepackage{amssymb}
\usepackage{comment}
\usepackage{amsmath}
\usepackage{braket}
\usepackage{mathrsfs}
\usepackage{mathtools}
\usepackage{array}
\usepackage{fancyhdr}
\usepackage[dvipdfmx]{graphicx}
\usepackage{color}
\usepackage{cite}
\usepackage{bm}
\usepackage{url}
\usepackage{fancyhdr}
\usepackage{booktabs}
\usepackage[colorlinks,citecolor=blue]{hyperref}
\usepackage[hang,small,bf]{caption}
\usepackage[subrefformat=parens]{subcaption}
\renewcommand{\thefootnote}{\#\arabic{footnote}}

\renewcommand{\thefootnote}{\fnsymbol{footnote}}
\def\thefootnote{\fnsymbol{footnote}}

\makeatletter

\@addtoreset{equation}{section}
\makeatother

\def\be{\begin{equation}}
\def\ee{\end{equation}}
\def\ben{\begin{eqnarray}}
\def\een{\end{eqnarray}}

\usepackage{scalerel}
\usepackage[usestackEOL]{stackengine}

\def\dashint{\,\ThisStyle{\ensurestackMath{%
            \stackinset{c}{.2\LMpt}{c}{.5\LMpt}{\SavedStyle-}{\SavedStyle\phantom{\int}}}%
        \setbox0=\hbox{$\SavedStyle\int\,$}\kern-\wd0}\int}

\begin{document}


\begin{center}

\vskip .75in

{\Large \bf Quantum description of gravitational waves generated by a classical source}

\vskip .75in

{\large
Felix Laga$\,^1$, Teruaki Suyama$\,^2$
}

\vskip 0.25in

{\em
$^{1}$Department of Physics, KU Leuven, 3001 Leuven, Belgium \\
$^{2}$Department of Physics, Institute of Science Tokyo, 2-12-1 Ookayama, Meguro-ku,
Tokyo 152-8551, Japan
}

\end{center}
\vskip .5in

\begin{abstract}
We investigate the quantum properties of gravitational waves (GWs) generated by a classical energy-momentum tensor. 
Treating the GW field as a quantum field coupled to a classical source, 
we evaluate the expectation value of the GW operator. 
We demonstrate that this expectation value exactly reproduces the classical retarded solution. 
Furthermore, we show that the mean and variance of the number of emitted gravitons
are equal. 
This suggests that the graviton emission is a Poisson process, as expected for a coherent state.
We establish a quantitative criterion for the validity of the classical wave description. 
By applying this criterion, we find that the classical approximation is remarkably accurate 
for astrophysical sources, but
laboratory-scale systems may reside in a regime where the discrete nature 
of graviton emission becomes significant.
\end{abstract}

\renewcommand{\thepage}{\arabic{page}}
\setcounter{page}{1}
\renewcommand{\thefootnote}{\#\arabic{footnote}}
\setcounter{footnote}{0}

\section{Introduction}
We are entering the golden era of gravitational-wave (GW) astronomy \cite{KAGRA:2021vkt}. 
Observations of GWs are becoming indispensable tools for a wide range of 
fields including cosmology, 
astrophysics, and gravitational physics. 
Typically, GWs emitted from astrophysical sources, such as binary systems, 
are calculated using classical general relativity \cite{Blanchet:2013haa}. 
Given the macroscopic scales of these systems, 
the classical treatment is generally considered a highly accurate approximation.
However, since the laws of nature are fundamentally governed by quantum mechanics, 
the physical processes underlying gravitational radiation must also possess a quantum mechanical description. 
Exploring this quantum-to-classical transition is not merely a theoretical exercise 
but is essential for a complete understanding of the nature of spacetime.

Recently, the quantum nature of GWs sourced by classical objects has attracted 
significant interest \cite{Kanno:2025how, Kanno:2025fpz, Das:2025kyn, Aoki:2026eos, Toccacelo:2026hcz}. 
In particular, it was argued in \cite{Kanno:2025how} that GWs emitted by binaries are 
described by coherent states at the level of the linearized gravity, 
and it was pointed out that the expectation value of the GW operator contains in-coming waves propagating from infinity toward the source. 
This result suggests a potential discrepancy between the quantum expectation value 
and the standard classical solution derived from the retarded Green's function.
In this short note, we revisit the quantum treatment of GWs, 
treating the energy-momentum tensor as a classical external source. 
Contrary to the claims above, 
we demonstrate that the expectation value of the quantum GW operator coincides 
exactly with the classical retarded solution. 
Furthermore, by evaluating the variance of the GW energy, 
we clarify the regime where the discrete nature of gravitons becomes prominent. 
We identify the specific conditions under which the classical wave picture fails 
and graviton emission becomes a rare, discrete Poisson process.
The results presented in this paper are fundamental and naturally expected within the framework of quantum field theory. 
While these properties might be considered well-established, 
recent discussions in the literature highlight that a rigorous re-examination of 
the quantum description of gravitational radiation remains both timely and essential. 
In this light, the present note aims to provide a necessary clarification and 
a solid theoretical foundation for the quantum-to-classical correspondence of GWs.

\section{Expectation value of GWs}
In this paper, we consider quantum GWs propagating on the Minkowski background
\be
ds^2=-dt^2+ \left( \delta_{ij}+h_{ij} \right) dx^i dx^j,
\ee
where $h_{ij}$ is given in the TT gauge ($\partial_i h_{ij}=\delta^{ij}h_{ij}=0$).
Expanding the Einstein-Hilbert action up to second order in $h_{ij}$, 
the kinetic term of $h_{ij}$ becomes
\be
S_g=\frac{1}{16\pi G}\int d^4x ~\sqrt{-g}R=\int d^4x ~\left(
\frac{1}{64\pi G}{\dot h_{ij}} {\dot h_{ij}}-\frac{1}{64\pi G} {\bm \nabla}h_{ij} \cdot {\bm \nabla} h_{ij} \right).
\ee
Then, $h_{ij}$ is coupled to the TT component of matter energy momentum tensor as
\be
S_{\rm int}= \int d^4x~ \frac{1}{2} T_{ij} h_{ij}.
\ee
In this paper, we treat $T_{ij}$ as a classical external force acting on 
the quantum field $h_{ij}$.

From the total action $S_g+S_{\rm int}$, the corresponding Hamiltonian becomes
\be
\label{Hamiltonian-gw}
H=\int d^3x \left( 16\pi G \pi_{ij} \pi_{ij}+\frac{1}{64\pi G}
{\bm \nabla}h_{ij} \cdot {\bm \nabla} h_{ij} -\frac{1}{2} T_{ij} h_{ij} \right),
\ee
where $\pi_{ij}$ is the conjugate momenta of $h_{ij}$.
We work in the Heisenberg picture where the operators ${\hat h}_{ij}$ and ${\hat \pi}_{ij}$ 
are promoted to time-dependent variables satisfying the commutation relations
\be
[ {\hat h}_{ij}(t,{\bm x}), {\hat \pi}_{k\ell} (t,{\bm y}) ]=\frac{i}{2}
\left( P_{ik} P_{j\ell}+P_{i\ell} P_{kj}-P_{ij}P_{k\ell} \right)
\delta ({\bm x}-{\bm y}) \times {\bm 1},
\ee
where $P_{ij}=\delta_{ij}-\frac{\partial_i \partial_j}{\triangle}$ is the projection operator and ${\bm 1}$ is an identity operator.
The other relations are
$[ {\hat h}_{ij}(t,{\bm x}), {\hat h}_{k\ell} (t,{\bm y}) ]=
[ {\hat \pi}_{ij}(t,{\bm x}), {\hat \pi}_{k\ell} (t,{\bm y}) ]=0$.
The corresponding Heisenberg equations are given by
\begin{align}
&\frac{\partial}{\partial t}{\hat h}_{ij}=i[{\hat H},{\hat h}_{ij}]
=32\pi G {\hat \pi}_{ij}, \\
&\frac{\partial}{\partial t}{\hat \pi}_{ij}=i[{\hat H},{\hat \pi}_{ij}]
=\frac{1}{32\pi G} \triangle {\hat h}_{ij}+\frac{1}{2}T_{ij} \times {\bm 1}.
\end{align}
From these equations, we obtain
\be
\left( -\frac{\partial^2}{\partial t^2} +\triangle \right) {\hat h}_{ij}
=-16\pi G T_{ij} \times {\bm 1}.
\ee
The solution of this equation satisfying the commutation relations is 
the inhomogeneous solution plus the homogeneous solution
\be
\label{sol-Heisenberg-eq}
{\hat h}_{ij}(t,{\bm x})=h^{\rm (cl)}_{ij}(t,{\bm x}) \times {\bm 1}
+{\hat h}^{\rm (free)}_{ij} (t,{\bm x}).
\ee
Here $h^{\rm (cl)}_{ij}(t,{\bm x})$, which is a c-number, is a solution
of the classical linearized Einstein equation
\be
\left( -\frac{\partial^2}{\partial t^2} +\triangle \right) h^{\rm (cl)}_{ij}
=-16\pi G T_{ij},
\ee
and ${\hat h}^{\rm (free)}_{ij} (t,{\bm x})$ is the free-field operator:
\be
\label{free-field}
{\hat h}^{\rm (free)}_{ij} (t,{\bm x})=\sqrt{32\pi G}
\int \frac{d^3p}{{(2\pi)}^{\frac{3}{2}} \sqrt{2p}} \sum_{\lambda=L,R}
\left( e^{i{\bm p}\cdot x-ip t} e_{ij}({\bm p},\lambda) 
{\hat a}({\bm p},\lambda)+
e^{-i{\bm p}\cdot x+ip t} e^*_{ij}({\bm p},\lambda) 
{\hat a}^\dagger ({\bm p},\lambda)\right).
\ee
The index $\lambda$ labels the polarization (either left-polarization
or right-polarization) and
$e_{ij}({\bm p},\lambda)$ is the polarization tensor satisfying the
normalization condition:
$e_{ij}({\bm p},\lambda) e^*_{ij} ({\bm p},\sigma)=\delta_{\lambda \sigma}$.
${\hat a}({\bm p},\lambda)$ and ${\hat a}^\dagger ({\bm p},\lambda)$
are annihilation and creation operators satisfying
\be
[ {\hat a}({\bm p},\lambda), {\hat a}^\dagger ({\bm q},\sigma)]
=\delta ({\bm p}-{\bm q}) \delta_{\lambda \sigma} \times {\bm 1}.
\ee
Using the fact that the retarded Green function is given by 
$G_{\rm ret}(x,x')=\frac{\Theta (t-t')}{4\pi |{\bm x}-{\bm x'}|}
\delta (t-t'-|{\bm x}-{\bm x'}|)$, 
$h^{\rm (cl)}_{ij}(t,{\bm x})$ can be expressed as
\be
\label{classical-sol}
h^{\rm (cl)}_{ij}(t,{\bm x})=4G \int d^3x'~\frac{1}{|{\bm x}-{\bm x'}|}
T_{ij} (t-|{\bm x}-{\bm x'}|,{\bm x'}).
\ee
In the far-field limit, assuming that the motion of the source is slow,
this solution becomes
\be
\label{classical-sol-2}
h^{\rm (cl)}_{ij}(t,{\bm x})=\frac{4G}{r} \int d^3x'~T_{ij} (t-r,{\bm x'}),
\ee
where $r=|{\bm x}|$.
This can be expressed by the quadrupole formula as
\be
\label{hcl-quadrupole-formula}
h^{\rm (cl)}_{ij}(t,{\bm x})=\frac{2G}{r} \left( {\cal P}_{ia} {\cal P}_{jb}
-\frac{1}{2} {\cal P}_{ij} {\cal P}_{ab} \right) 
{\ddot Q}_{ab} (t_r).
\ee
Here ${\cal P}_{ij}=\delta_{ij}-n_i n_j$ is the operator that projects onto a plane
normal to unit vector ${\bm n}=\frac{\bm x}{r}$ and $Q_{ij}$ is the reduced 
quadrupole defined by
\be
Q_{ij}=\int d^3x ~\rho (t,{\bm x}) x^i x^j-\frac{1}{3}\delta_{ij} \int d^3x
~\rho(t,{\bm x}) x^2.
\ee

Now suppose that $T_{ij}$ was inactive in the distant past and turned on at some point and that the state was initially in the vacuum state $|0\rangle$ defined by the condition
${\hat a}({\bm p},\lambda) |0\rangle =0$.
Then, using Eq.~(\ref{sol-Heisenberg-eq}),
the expectation value of ${\hat h}_{ij}$ at the spacetime point $(t,{\bm x})$
is evaluated as
\be
\label{vac-expec}
\langle 0| {\hat h}_{ij}(t,{\bm x}) |0\rangle =h^{\rm (cl)}_{ij}(t,{\bm x}).
\ee
Thus, the expectation value exactly coincides with the classical solution.
In Appendix \ref{appendix-1}, based on the interaction picture, 
we reproduce this result with RHS being given by Eq.~(\ref{hcl-quadrupole-formula}).

Before closing this section, as an example, 
we directly evaluate Eq.~(\ref{classical-sol-2}) for a binary consisting of two point particles. 
The action for the point particles in the non-relativistic regime is given by
\be
S_p=\int dt \left( \frac{m_1}{2}{\dot {\bm x_1}}^2+\frac{m_2}{2}{\dot {\bm x_2}}^2-
V({\bm x}_1, {\bm x}_2) \right),
\ee
where $V({\bm x}_1, {\bm x}_2)$ is the binding energy. 
In the presence of GWs, not $|{\bm x}|$ but $g_{ij} x^i x^j$ represents the proper distance.
New coordinates ${\bm X}$ defined by 
\be
\label{change-coordinates}
X^i=x^i+\frac{1}{2}h_{ij}x^j
\ee
satisfies $\delta_{ij}X^i X^j=g_{ij}x^i x^j$.
This indicates that ${\bm X}$ directly provides the proper distance and 
the argument of $V$ should now be ${\bm X}$.
Thus, in the presence of GWs, the action becomes
\be
S_p=\int dt \left( \frac{m_1}{2}g_{ij} {\dot x_1}^i {\dot x_1}^j
+\frac{m_2}{2}g_{ij} {\dot x_2}^i {\dot x_2}^j
-V({\bm X}_1, {\bm X}_2) \right).
\ee
Substituting Eq.~(\ref{change-coordinates}) and $g_{ij}=\delta_{ij}+h_{ij}$
and expanding the action up to first order in $h_{ij}$, we obtain 
\be
S_p \approx \int dt \left( \frac{m_1}{2}{\dot {\bm x_1}}^2
+\frac{m_2}{2}{\dot {\bm x_2}}^2-V ({\bm x_1},{\bm x_2})
+\frac{m_1}{2} h_{ij} {\dot x_1}^i {\dot x_1}^j
+\frac{m_2}{2} h_{ij} {\dot x_2}^i {\dot x_2}^j
-\frac{1}{2} h_{ij} \frac{\partial V}{\partial x_1^i} x_1^j 
-\frac{1}{2} h_{ij} \frac{\partial V}{\partial x_2^i} x_2^j\right).
\ee
Using this expression, the energy-momentum tensor becomes
\be
\label{EMT-circular-binary}
T^{ij} (t,{\bm x})=m_1 \left( v_1^i v_1^j+\frac{1}{2}(x_1^i a_1^j+x_1^j a_1^i) \right) \delta ({\bm x}-{\bm x_1}(t))+(1 \to 2),
\ee
where ${\bm v}={\dot {\bm x}}, {\bm a}={\ddot {\bm x}}$ and we have used 
the Newton's equation of motion $m{\bm a}=-{\bm \nabla} V$.

For simplicity, let us assume a circular binary confined on the $x$--$y$ plane:
\begin{align}
&x_1(t)=\frac{m_2}{M}R \cos (\Omega t),~~~~~y_1(t)=\frac{m_2}{M}R \sin (\Omega t), \\
&x_2(t)=-\frac{m_1}{M}R \cos (\Omega t),~~~~~y_2 (t)=-\frac{m_1}{M}R \sin (\Omega t),
\end{align}
where $M=m_1+m_2,~\mu=\frac{m_1 m_2}{M}$ and $R$ is the distance between the point particles.
Then, at large distance from the source ($r=|{\bm x}|\gg R$), 
Eqs.~(\ref{classical-sol-2}) and (\ref{vac-expec}) yield
\begin{align}
&\langle 0| {\hat h}_{xx} |0 \rangle =h^{\rm (cl)}_{xx} = -\frac{4G\mu R^2\Omega^2}{r} \cos (2\Omega t_r), \label{bexp-xx}\\
&\langle 0|{\hat h}_{xy} |0\rangle =
h^{\rm (cl)}_{xy} = -\frac{4G\mu R^2\Omega^2}{r} \sin (2\Omega t_r), \\
&\langle 0| {\hat h}_{yy} |0\rangle =h^{\rm (cl)}_{yy} = \frac{4G\mu R^2\Omega^2}{r} \cos (2\Omega t_r), \label{bexp-yy}
\end{align}
where $t_r=t-r$ is the retarded time.
These expressions of the expectation values match exactly those obtained 
by using the quadrupole formula in classical theory.

\section{Quantum fluctuations and graviton statistics}
Eq.~(\ref{classical-sol}) shows that the expectation value of quantum GWs always 
coincides with its classical counterpart. 
However, this does not necessarily mean that quantum GWs always reside in the classical regime. For a classical treatment to provide an approximately accurate description, 
quantum fluctuations must be sufficiently suppressed relative to the classical values. 
To examine this point in more detail, we consider the energy of the GWs emitted from the classical and localized system undergoing periodic non-relativistic motion. 
Given that the energy takes discrete values (i.e., integer times $\hbar \omega$) 
when GWs are quantized, 
quantum aspects are expected to manifest more prominently in the energy than in the GW amplitude.

Based on the GW luminosity expression in GR, the corresponding GW luminosity operator is given by
\be
\label{def-luminosity}
{\hat L}=-\frac{1}{32\pi G}\int_{S_2} dS~n^k \partial_t {\hat h}_{ij} ~{\hat h}_{ij,k},
\ee
where $S_2$ is a sphere of radius $r$ with the GW source located at the center
\footnote{To ensure gauge invariance, 
Eq.~(\ref{def-luminosity}) should be averaged over a period. 
However, since we will ultimately evaluate the time integral of Eq.~(\ref{def-luminosity}), 
we omit the averaging process here to avoid redundancy.}. 
$r$ is taken to be much larger than the size
of the GW source.
Substituting the solution (\ref{sol-Heisenberg-eq}), we obtain
\be
\label{operator-luminosity}
{\hat L}=L^{\rm (cl)}\times {\bm 1} -\frac{1}{32\pi G}\int_{S_2} dS~n^k \partial_t{\hat h}^{\rm (free)}_{ij}  
h^{\rm (cl)}_{ij,k}
-\frac{1}{32\pi G}\int_{S_2} dS~n^k \partial_t h^{\rm (cl)}_{ij}  
{\hat h}^{\rm (free)}_{ij,k}.
\ee
Here $L^{\rm (cl)}$, which contains only $h^{\rm (cl)}$, 
is the GW luminosity in classical theory and
the term containing only ${\hat h}^{\rm (free)}_{ij}$ is omitted because
it is irrelevant for our purpose.

Using Eqs.~(\ref{free-field}) and (\ref{hcl-quadrupole-formula}), 
the second term of Eq.~(\ref{operator-luminosity}) becomes
\begin{align}
&-\frac{1}{32\pi G}\int_{S_2} dS~n^k \partial_t{\hat h}^{\rm (free)}_{ij}  
h^{\rm (cl)}_{ij,k} \nonumber \\
&=-ir \sqrt{\frac{G}{8\pi}} \int d\Omega \int \frac{d^3p}{{(2\pi)}^\frac{3}{2}}
\sqrt{\frac{p}{2}} \sum_\lambda \left( e^{ir {\bm p}\cdot {\bm n}-ipt} e_{ij}({\bm p},\lambda)
{\hat a}({\bm p},\lambda)-e^{-ir {\bm p}\cdot {\bm n}+ipt} e^*_{ij}({\bm p},\lambda)
{\hat a}^\dagger ({\bm p},\lambda) \right) \nonumber \\
&\hspace{10mm}\times \left( {\cal P}_{ia} {\cal P}_{jb}
-\frac{1}{2} {\cal P}_{ij} {\cal P}_{ab} \right) {\dddot Q}_{ab} (t_r),
\end{align}
where $\int d\Omega$ is integration over ${\bm n}$.
Using 
\be
\int d\Omega~ n_{i_1} \cdots n_{i_N}e^{ir {\bm p}\cdot {\bm n}}
=\frac{4\pi}{{(ir)}^N}\frac{\partial}{\partial p_{i_1}}\cdots
\frac{\partial}{\partial p_{i_N}} j_0 (pr),
\ee
we have
\begin{align}
&\int d\Omega ~n_i n_j e^{ir {\bm p}\cdot {\bm n}}=4\pi j_0 (pr) {\hat p}_i {\hat p}_j+{\cal O}
\left( \frac{1}{r^2} \right), \label{int-nn} \\
&\int d\Omega ~n_i n_j n_a n_b e^{ir {\bm p}\cdot {\bm n}}=4\pi j_0 (pr) 
{\hat p}_i {\hat p}_j {\hat p}_a {\hat p}_b +{\cal O}
\left( \frac{1}{r^2} \right),
\end{align}
where ${\hat {\bm p}}=\frac{\bm p}{p}$.
Then, using $e_{ij}({\bm p},\lambda){\hat p}_j=0$, we obtain
\begin{align}
&-\frac{1}{32\pi G}\int_{S_2} dS~n^k \partial_t{\hat h}^{\rm (free)}_{ij}  
h^{\rm (cl)}_{ij,k} \nonumber \\
&=-i \sqrt{2\pi G}  \int \frac{d^3p}{{(2\pi)}^\frac{3}{2}}
\frac{\sin (pr)}{\sqrt{2p}} \sum_\lambda 
\left( e^{-ipt} e_{ij}({\bm p},\lambda)
{\hat a}({\bm p},\lambda)-e^{ipt} e^*_{ij}({\bm p},\lambda)
{\hat a}^\dagger ({\bm p},\lambda) \right) {\dddot Q}_{ij}(t_r).
\end{align}
In a similar manner, the third term of Eq.~(\ref{operator-luminosity}) becomes
\begin{align}
&-\frac{1}{32\pi G}\int_{S_2} dS~n^k \partial_t h^{\rm (cl)}_{ij}  
{\hat h}^{\rm (free)}_{ij,k} \nonumber \\
&=-\sqrt{2\pi G}  \int \frac{d^3p}{{(2\pi)}^\frac{3}{2}}
\frac{\cos (pr)}{\sqrt{2p}} \sum_\lambda 
\left( e^{-ipt} e_{ij}({\bm p},\lambda)
{\hat a}({\bm p},\lambda)+e^{ipt} e^*_{ij}({\bm p},\lambda)
{\hat a}^\dagger ({\bm p},\lambda) \right) {\dddot Q}_{ij}(t_r).
\end{align}
Putting these results together, we obtain
\be
\label{operator-luminosity-2}
{\hat L}=L^{\rm (cl)}\times {\bm 1}+
\int \frac{d^3p}{{(2\pi)}^\frac{3}{2}} \sum_\lambda
\left( e^{ip(r-t)}f_{{\bm p},\lambda} (t_r) {\hat a}({\bm p},\lambda) 
+e^{-ip(r-t)}f^*_{{\bm p},\lambda} (t_r) {\hat a}^\dagger ({\bm p},\lambda) \right),
\ee
where we have introduced 
\be
f_{{\bm p},\lambda} (t_r) =-\sqrt{\frac{\pi G}{p}} e_{ij}({\bm p},\lambda)
{\dddot Q}_{ij}(t_r).
\ee

Given the operator ${\hat L}$, an operator ${\hat E}$ corresponding to the energy of
GWs emitted during $-\frac{T}{2} \le t \le \frac{T}{2}$ is given by
\be
{\hat E}=\int_{-\frac{T}{2}}^{\frac{T}{2}} {\hat L}~dt.
\ee
Here we take $T$ to be larger than the period of $Q_{ij}$
(=period of GWs) to grant ${\hat E}$ the gauge invariance.
Using Eq.~(\ref{operator-luminosity-2}), ${\hat E}$ becomes
\be
\label{operator-energy}
{\hat E}=E^{\rm (cl)}\times {\bm 1}+
\int \frac{d^3p}{{(2\pi)}^\frac{3}{2}} \sum_\lambda
\left( g_{{\bm p},\lambda} {\hat a}({\bm p},\lambda) 
+g^*_{{\bm p},\lambda} {\hat a}^\dagger ({\bm p},\lambda) \right),
\ee
where 
\be
g_{{\bm p},\lambda}  =\int_{-\frac{T}{2}}^{\frac{T}{2}} e^{-ipt_r} f_{{\bm p},\lambda}(t_r) dt.
\ee

Now, we are ready to compute expectation value and variance of ${\hat E}$. 
In what follows, we suppose that each component $Q_{ij}(t)$ is proportional to sinusoidal functions as
\be
Q_{ij}(t)=A_{ij} \cos (\omega_0 t+\beta_{ij}),
\ee
where $\beta_{ij}$ is phase.
Then, the classical part becomes
\be
E^{\rm (cl)}=\int_{-\frac{T}{2}}^{\frac{T}{2}} \frac{G}{5} {\dddot Q}_{ij}{\dddot Q}_{ij}~dt
=\frac{G}{10}\omega_0^6 A_{ij}A_{ij}T.
\ee
From Eq.~(\ref{operator-energy}), the expectation value of ${\hat E}$ of the state
which was in the vacuum state in the distant past becomes
\be
\label{mean-energy}
\langle 0|{\hat E} |0\rangle =E^{\rm (cl)}=\frac{G}{10}\omega_0^6 A_{ij}A_{ij}T.
\ee
Thus, as it is the case for the GW amplitude $h_{ij}$, the expectation value of the
GW energy coincides with the classical one.

The variance of ${\hat E}$ is defined by
\be
{(\Delta E)}^2=\langle 0| {({\hat E}-E^{\rm (cl)}\times {\bm 1})}^2 |0\rangle.
\ee
Substituting Eq.~(\ref{operator-energy}), we obtain
\be
{(\Delta E)}^2=\int \frac{d^3p}{{(2\pi)}^3} \sum_\lambda
{|g_{{\bm p},\lambda}|}^2.
\ee
From the definition of $g_{{\bm p},\lambda}$, we have
\begin{align}
g_{{\bm p},\lambda}&=-\sqrt{\frac{\pi G}{p}} \omega_0^3 e_{ij}({\bm p},\lambda)
A_{ij} \int_{-\frac{T}{2}}^{\frac{T}{2}} e^{-i p (t-r)} \sin (\omega_0 (t-r)+\beta_{ij}) dt \nonumber \\
&\approx \pi i \sqrt{\frac{\pi G}{p}} \omega_0^3 e_{ij}({\bm p},\lambda)
A_{ij} \delta (p-\omega_0) e^{i\beta_{ij}}, 
\end{align}
where we have used an approximation $\int_{-\frac{T}{2}}^{\frac{T}{2}} 
e^{i p t}dt \approx 2\pi \delta (p)$ for large $T$.
Using the formula
\be
\label{formula-ee}
\sum_\lambda e_{ij}({\bm p},\lambda) e^*_{ab} ({\bm p},\lambda)
=\frac{1}{2} P'_{ia} P'_{jb}+\frac{1}{2}P'_{ib}P'_{ja}-\frac{1}{2}P'_{ij}P'_{ab},~~~~~P'_{ij}=\delta_{ij}-{\hat p}_i {\hat p}_j, 
\ee
we obtain
\be
{(\Delta E)}^2=\frac{G\omega_0^6 T}{32\pi} A_{ij}A_{ab}
\int dp~p \delta (p-\omega_0) \int d\Omega 
(P'_{ia} P'_{jb}+P'_{ib}P'_{ja}-P'_{ij}P'_{ab}).
\ee
Here, one of the delta functions from the squared delta function $\delta^2 (p-\omega_0)$
has been replaced using the relation $\delta (0)=\frac{T}{2\pi}$, 
which is valid for $p=\omega_0$.
Additionally, $\int d\Omega$ denotes the integration over the unit sphere spanned by
${\hat {\bm p}}$.
Using the formula
\be
\int d\Omega 
(P'_{ia} P'_{jb}+P'_{ib}P'_{ja}-P'_{ij}P'_{ab})=
\frac{8\pi}{15} \left( 3\delta_{ia} \delta_{jb}+3\delta_{ib}\delta_{ja}
-2\delta_{ij} \delta_{ab} \right),
\ee
we arrive at the final expression
\be
\label{variance-energy}
{(\Delta E)}^2=\frac{G T}{10}\omega_0^7 A_{ij}A_{ij}.
\ee

In the quantum mechanical description, monochromatic GWs with frequency $\omega_0$
are interpreted as a collection of gravitons, each having energy $\omega_0$.
Thus, from Eq.~(\ref{mean-energy}),
the mean number of gravitons $N$ emitted from the source during time $T$
is given by
\be
\label{number-gravitons}
N=\frac{GT}{10}\omega_0^5 A_{ij}A_{ij}.
\ee
Thus, the source emits one graviton on average every $\tau$,
where $\tau$ is given by
\be
\tau=\frac{T}{N}=\frac{10}{G\omega_0^5 A_{ij}A_{ij}}.
\ee
From Eq.~(\ref{variance-energy}), the variance of the number fluctuation
$\delta N$ is given by
\be
{(\delta N)}^2=\frac{GT}{10}\omega_0^5 A_{ij}A_{ij}=N.
\ee
This result, $\delta N=\sqrt{N}$, indicates that graviton emission is a Poisson process. 
This is consistent with the fact that the occupation number of a coherent 
state follows a Poisson distribution \cite{Glauber:1963tx}, 
as is the case in our setup \cite{Kanno:2025how}.
Given that the wave picture is valid only for $T \gtrsim \omega_0^{-1}$, 
we can conclude that GWs deviate significantly from classical behavior 
if not even a single graviton is emitted during a single oscillation period: 
$\tau > \omega_0^{-1}$.

We evaluate $N_g$, the number of gravitons emitted during one oscillation period, for several examples found in standard textbooks. 
Table~\ref{tab:comparison} lists the values of $N_g$ obtained using Eq.~(\ref{number-gravitons}) for $T=2\pi/\omega_0$.
In MTW \cite{MTW}, a massive steel beam with a radius of $1~{\rm m}$, 
a length of $20~{\rm m}$, a mass of $490$ tons, 
and a tensile strength of $3\times 10^9~{\rm dyne/cm^2}$ is considered as a source of GWs. 
The beam is assumed to rotate about its center with an angular velocity 
($\omega = 28~{\rm rad/s}$) limited by the balance between centrifugal force and tensile strength. 
For such a source, we obtain $N_g \approx 420$. 
Since this is much larger than unity, a classical description of the 
GW emission process is appropriate.
In Schutz \cite{Schutz2009}, two identical masses (each $10^3~{\rm kg}$) connected by a massless spring undergo oscillations 
with an amplitude of $10^{-4}~{\rm m}$ and an angular frequency of $\omega=10^4~{\rm s^{-1}}$ about equilibrium positions $1~{\rm m}$ apart. 
In this case, we find that $N_g$ is approximately $9\times 10^{-6}$. 
Thus, the classical picture is strictly invalid, and gravitons are 
rarely emitted during a single period of motion.
In Creighton \& Anderson \cite{CreightonAnderson2011}, 
a rotating bar with a mass of $1~{\rm kg}$ and a length of $1~{\rm m}$ 
spinning at $\omega=1~{\rm rad/s}$ is considered (see also Wald \cite{Wald1984}). For this system, we find $N_g \approx 1.4\times 10^{-20}$, 
implying that not even a single graviton would be emitted over the age of the universe.
Finally, in Weinberg \cite{Weinberg1972}, 
the orbital motion of Jupiter is considered. 
In this case, we find $N_g \approx 7\times 10^{53}$. 
This is a vast number, confirming that the classical description 
is an excellent approximation.

\begin{table}[tp] 
  \centering 
  \begin{tabular}{| >{\centering\arraybackslash}m{3.5cm} | >{\centering\arraybackslash}m{6cm} | >{\centering\arraybackslash}m{2.5cm} |} \hline
    \textbf{Textbook} & \textbf{System} & \boldmath $N_g$ \\ \hline \hline
    MTW \cite{MTW} & Rotating steel beam with maximal rotation speed & $4\times 10^2$ \\ \hline
    Schutz \cite{Schutz2009} & Two mass points attached to an oscillating spring & $9\times 10^{-6}$ \\ \hline
    Creighton \& Anderson \cite{CreightonAnderson2011} & Rotating bar & $1.4\times 10^{-20}$ \\ \hline
    Weinberg \cite{Weinberg1972} & Jupiter orbiting around the sun & $7\times 10^{53}$ \\ \hline
  \end{tabular}
  \caption{The number of gravitons emitted per oscillation period ($N_g$) for various physical systems introduced in standard textbooks.} 
  \label{tab:comparison} 
\end{table}

\section{Summary}
In this paper, we have investigated the quantum properties of GWs sourced 
by a classical energy-momentum tensor. 
Our analysis confirms that the expectation value of the quantum GW operator 
exactly coincides with the classical solution derived from the retarded Green's function. 
This consistency holds in both the Heisenberg and interaction pictures, 
reaffirming the robustness of the correspondence between the quantum field-theoretic description and the standard classical framework of GR.
Specifically, we have clarified the origin of the in-coming waves reported in the recent study \cite{Kanno:2025how}. 
This resolution proves that the alleged difference between the quantum expectation value 
and the classical retarded solution is not a physical effect 
but rather a consequence of the mathematical prescription used in the evaluation.
Furthermore, we extended our analysis beyond the expectation value 
by evaluating the variance of the GW energy operator to probe the statistical nature of gravitational radiation. 
The fact that the fluctuation in the number of emitted gravitons 
follows the relation $\delta N = \sqrt{N}$ indicates that graviton emission 
from a classical source is a Poisson process. 
Based on these results, we identified a quantitative criterion for the validity 
of the classical wave description, defined by the number of gravitons emitted per oscillation period. 
For macroscopic astrophysical systems, such as the orbital motion of Jupiter, the classical approximation is extremely accurate. 
In contrast, for laboratory-scale systems like mechanical oscillators or rotating bars, 
the classical wave picture fails, and the radiation process is dominated 
by the discrete and rare emission of individual gravitons.
The quantitative boundaries established in this work provide a solid theoretical 
foundation for distinguishing between the classical wave regime 
and the quantum regime where the discrete nature of gravity becomes prominent.

\section*{Acknowledgments}
This work was supported by JSPS KAKENHI Grant Number JP23K03411 (TS).

\appendix
\section{Derivation of Eq.~(\ref{vac-expec}) in the interaction picture}
\label{appendix-1}
It may be useful to derive Eq.~(\ref{vac-expec}),
which has been derived based on the Heisenberg picture in the main text,
based on the interaction picture in this Appendix.

Our starting point is the Hamiltonian given by Eq.~(\ref{Hamiltonian-gw}).
We will treat the last term $-\frac{1}{2} T_{ij}h_{ij}$ as the interaction term.
In the interaction picture, the time evolution of operators is governed by the 
free part of the Hamiltonian, while that of states is driven by the interaction part.
Thus, the operator of $h_{ij}$ in the interaction picture is nothing but 
${\hat h}^{\rm (free)}_{ij} (t,{\bm x})$ given by Eq.~(\ref{free-field}).
The state is evolved by the unitary operator ${\hat U}_I(t)$ given by
\be
{\hat U}_I(t)={\cal T} \exp \left( -i \int_{-\infty}^t dt'~{\hat H}_I (t') \right),
\ee
where ${\cal T}$ is the time-ordering operator and 
${\hat H}_I =-\int d^3x \frac{1}{2}T_{ij}{\hat h}^{\rm (free)}_{ij}$.
Using Eq.~(\ref{free-field}), we have
\be
\int_{-\infty}^t d\tau \int_{-\infty}^\tau d\tau'~[ {\hat H}_I (\tau), {\hat H}_I (\tau')]
=2i \theta,
\ee
where $\theta$, which is real, is defined by
\be
\theta =8\pi G \int_{-\infty}^t d\tau \int_{-\infty}^\tau d\tau'~
\int \frac{d^3p}{{(2\pi)}^3 2p} \Im \left( \sum_\lambda {\tilde T}_\lambda (\tau,{\bm p})
e^{-ip \tau} {\tilde T}^*_\lambda (\tau',{\bm p}) e^{ip\tau'} \right),
\ee
where ${\tilde T}_{ij}(\tau, {\bm p}) \equiv \int d^3x ~e^{i{\bm p}\cdot {\bm x}} T_{ij}(\tau,{\bm x})$
and ${\tilde T}_\lambda (\tau,{\bm p}) \equiv {\tilde T}_{ij}(\tau,{\bm p}) e_{ij} ({\bm p},\lambda)$.
Then, we have
\be
{\hat U}_I (t)=e^{-i\theta} {\hat D}(\alpha),
\ee
where ${\hat D}(\alpha)$ is the so-called displacement operator
\be
{\hat D}(\alpha)=\exp  \bigg[ \sum_\lambda \int d^3p \left( \alpha ( {\bm p},\lambda) 
{\hat a}^\dagger ({\bm p}, \lambda)-\alpha^* ({\bm p},\lambda) 
{\hat a} ({\bm p}, \lambda) \right) \bigg],
\ee
and $\alpha ({\bm p}, \lambda)$ is the coherent state parameter defined by
\be
\alpha ({\bm p},\lambda )=i \sqrt{\frac{G}{2\pi^2 p}}\int_{-\infty}^t
dt' ~\int d^3x~ T_{ij}(t',{\bm x}) e^*_{ij}({\bm p},\lambda) 
e^{-i {\bm p}\cdot {\bm x}+ipt'}.
\ee
Starting from the vacuum state $|0\rangle$ in the distant past,
the state at time $t$ in the interaction picture is given by
${\hat U}_I(t) |0\rangle =e^{-i\theta} {\hat D}(\alpha) |0\rangle$.
Using the formula ${\hat D}^\dagger (\alpha) {\hat a}({\bm p},\lambda){\hat D}(\alpha)={\hat a}({\bm p},\lambda)+\alpha ({\bm p},\lambda) \times {\bm 1}$,
the expectation value of the operator ${\hat h}_{ij}$ becomes
\begin{align}
&\langle 0 | {\hat U}_I^\dagger (t) {\hat h}^{\rm (free)}_{ij} (t,{\bm x}) 
{\hat U}_I(t) |0\rangle \nonumber \\
&=\sqrt{32\pi G}\sum_\lambda \int \frac{d^3p}{{(2\pi)}^\frac{3}{2} \sqrt{2p}}
\bigg[ e_{ij}({\bm p},\lambda) \alpha ({\bm p},\lambda)e^{i{\bm p}\cdot {\bm x}-ipt} +e^*_{ij}({\bm p},\lambda) \alpha^* ({\bm p},\lambda)e^{-i{\bm p}\cdot {\bm x}+ipt} \bigg] \nonumber \\
&=8 \pi iG\sum_\lambda \int d^3 x'\int_{-\infty}^t dt' \int 
\frac{d^3 p}{{(2\pi)}^3 p} e_{ij}({\bm p},\lambda) e^*_{ab} ({\bm p},\lambda)
e^{i {\bm p}\cdot ({\bm x}-{\bm x}')-ip (t-t')} T_{ab} (t',{\bm x}')
+{\rm c.c.}, \nonumber
\end{align}
where ${\rm c.c.}$ means complex conjugate.
Using the formula (\ref{formula-ee}) and assuming the far wave-zone
like Eq.~(\ref{int-nn}), we obtain
\begin{align}
\label{appendix-exp-hij}
&\langle 0 | {\hat U}_I^\dagger (t) {\hat h}^{\rm (free)}_{ij} (t,{\bm x}) 
{\hat U}_I(t) |0\rangle \nonumber \\
&=\frac{2iG}{\pi r} \left( {\cal P}_{ia} {\cal P}_{jb}+ {\cal P}_{ib} {\cal P}_{ja}
-{\cal P}_{ij} {\cal P}_{ab} \right)
\int d^3 x' \int_{-\infty}^tdt' T_{ab}(t',{\bm x}')
\int_0^\infty dp~\sin (pr) e^{-ip (t-t')}+{\rm c.c.} \nonumber \\
&=\frac{2G}{r}\left( {\cal P}_{ia} {\cal P}_{jb}+ {\cal P}_{ib} {\cal P}_{ja}
-{\cal P}_{ij} {\cal P}_{ab} \right) \int d^3x' ~T_{ab} (t-r,{\bm x'}).
\end{align}
Using the formula
\be
\int d^3x' ~T_{ab} (t-r,{\bm x'})
=\frac{1}{2} {\ddot I}_{ij} (t-r),~~~~~~I_{ij}=\int d^3 x'~x^i x^j
\rho (t-r,{\bm x}'),
\ee
we finally obtain
\be
\langle 0 | {\hat U}_I^\dagger (t) {\hat h}^{\rm (free)}_{ij} (t,{\bm x}) 
{\hat U}_I(t) |0\rangle
=\frac{2G}{r} \left( {\cal P}_{ia} {\cal P}_{jb}
-\frac{1}{2} {\cal P}_{ij} {\cal P}_{ab} \right) 
{\ddot Q}_{ab} (t_r).
\ee
This is nothing but the quadrupole formula of classical GWs.
In \cite{Kanno:2025how}, it was shown that the expectation value
of $h_{ij}$ in the interaction picture contains not only out-going waves
but also in-coming waves with the same amplitude as the out-going waves.
In essence, the discrepancy can be understood from the following double integral 
\be
I=\int_0^\infty dp \sin (pr) \int_{-\infty}^t dt' 
\big[ \sin (pt-pt'+\Omega t') +\sin (pt-pt'-\Omega t') \big],
\ee
which appears in the computation of the expectation value of GWs ($\Omega>0, r>0$).
In \cite{Kanno:2025how}, the integral over $t'$ was first performed.
Because the first term in the square brackets oscillates except for $p=\Omega$,
it was replaced with $\pi \sin (\Omega t) \delta (p-\Omega)$. 
The second term was ignored because it oscillates for any value of $p$.
With this prescription, the integral $I$ becomes
\be
\label{appendix-def-I}
I \simeq \pi \sin (\Omega r) \sin (\Omega t)
=\frac{\pi}{2} \cos (\Omega (r-t)) -\frac{\pi}{2} \cos (\Omega (r+t)),
\ee
and the second term represents the in-coming waves.

On the other hand, by using the formula 
\be
\int_{-\infty}^0 e^{i\omega \tau}d\tau=\frac{1}{i} PV \frac{1}{\omega}+\pi \delta (\omega),
\ee
where $PV$ stands for the Cauchy principal value,
we can exactly evaluate the integral of the first term in $I$ as
\be
\int_{-\infty}^t dt'~\sin (pt-pt'+\Omega t')=\cos (\Omega t)
PV \frac{1}{p-\Omega}+\pi \sin (\Omega t) \delta (p-\Omega).
\ee
Then, the integral $I$ becomes
\be
I=\pi \sin (\Omega r) \sin (\Omega t)+\cos (\Omega t) \int_0^\infty dp
~\sin (pr) PV \frac{1}{p-\Omega}.
\ee
Using the formula
\be
\int_0^\infty dy~\sin (xy) PV \frac{1}{y-1}=-{\rm Ci}(x) \sin (x)+\frac{1}{2}
\cos (x) (\pi +2{\rm Si}(x)) =\pi \cos (x)+{\cal O}\left( \frac{1}{x}\right),
\nonumber
\ee
we finally obtain
\be
I=\pi \sin (\Omega r) \sin (\Omega t)+\pi \cos (\Omega t ) \cos (\Omega r)=
\pi \cos (\Omega (t-r)).
\ee
Thus, $I$ contains only out-going waves.

Alternatively, as we have done in obtaining Eq.~(\ref{appendix-exp-hij}),
performing the integration over $p$ first in Eq.~(\ref{appendix-def-I}) yields
\be
I=\pi \int_{-\infty}^t dt'~\big[ \delta (r-t+t')-\delta (r+t-t') \big]
\cos (\Omega t').
\ee
Then, given that $t'\le t$, the second term does not contribute and we 
end up with $I=\pi \cos (\Omega (t-r))$,
which contains only out-going waves.

\bibliography{draft}

\end{document}